\documentclass{article}
\newcommand{\be}{\begin{equation}}    % for lazy typers
\newcommand{\ee}{\end{equation}}
\newcommand{\ba}{\begin{eqnarray}}
\newcommand{\ea}{\end{eqnarray}}
\newcommand{\sizedef}{
        \headheight=0pt                               % zero space header
 	  \topmargin=-1.5cm \headsep=1.5cm              % put near top of page
        \oddsidemargin=-0.5cm \evensidemargin=-0.5cm  % adjust left
        \textheight=22truecm \textwidth=16.5truecm    % PLAIN text dimensions
 	  \setlength{\columnsep}{20pt}                  % default = 10pt
  }
 \usepackage{amsmath,amssymb}
\sizedef
\begin{document}
\bibliographystyle{prsty}

\title{INTEGRABLE HAMILTONIAN SYSTEMS WITH VECTOR POTENTIALS}

\author{ 
Giuseppe Pucacco\thanks{e-mail: pucacco@roma2.infn.it} \\ 
Dipartimento di Fisica -- Universit\`a di Roma ``Tor Vergata" \\ 
INFN -- Sezione di Roma II\\
Kjell Rosquist\thanks{e-mail: kr@physto.se} \\
Department of Physics -- Stockholm University} 
\date{}
\maketitle 
\vspace{2cm}
 
\begin{abstract}
We investigate integrable 2-dimensional 
Hamiltonian systems with scalar and vector potentials, 
admitting second invariants which are linear or quadratic in the momenta. In the case of a linear second invariant, we provide some examples of weakly-integrable systems. In the case of a quadratic second invariant, we recover the classical strongly-integrable systems in {\it Cartesian} and {\it polar} coordinates and provide some new examples of integrable systems in {\it parabolic} and {\it elliptical} coordinates. 
\end{abstract}
\vspace{2cm}

Submitted to {\it Journal of Math. Phys.} on May 26, 2004

\vfill\eject

\section{Introduction}
 The direct approach to investigate integrable Hamiltonian systems is 
a very classical subject \cite{darboux:inv, Whit}. It 
consists in determining the class of potentials supporting additional 
invariants within some specified family of phase-space functions. This method 
produced several interesting results in the eighties, as illustrated 
in the review by Hietarinta \cite{hiet}. Recently \cite{geom,max,kprs} 
the approach has been applied to treat in a unified 
way both invariants at fixed and arbitrary energy.   

Many results 
are known for natural reversible Hamiltonians. One of the reasons for 
this is that the search for additional invariants can be restricted 
to functions with a definite parity in the momenta. This property 
leads to a substantial reduction in the usually very complicated set 
of equations. Much less is known in the case of Hamiltonians with 
vector potentials. For a long time, the only systematic attempt to cope with this case was that of Dorizzi, Grammaticos, Ramani and 
Winternitz \cite{dgrw} providing a set of solutions in Cartesian coordinates. Recently, McSween and Winternitz \cite{msw} obtained some new solutions in polar coordinates and B\'erub\'e and Winternitz \cite{bw} extended the results to the corresponding quantum problem. In both works the authors also identify the subset of superintegrable systems. In an attempt to extend these results to include weak integrability, we have provided \cite{gp} a general solution for linear invariants and analyzed some new classes of weakly-integrable systems. 

The purpose of the present paper is to 
reinvestigate Hamiltonian systems with both scalar and vector potentials, 
trying to identify those admitting the 
existence of a second invariant which is a quadratic 
polynomial in the momenta. We state the general 
approach at arbitrary and fixed value of the first invariant (Jacobi constant) and show that, in the case of strong integrability, it is possible to get a general formal solution. This is valid for every standard coordinate systems which are the same as the separable ones in the purely scalar case. Therefore, in addition to the above mentioned Cartesian and polar case, solutions in parabolic and elliptical coordinates can be looked for. In all cases, the potentials are defined in terms of a pair of scalar functions for which we get the integrability conditions: solving them determines the vector potential, whereas the scalar potential is subject to an additional linear differential equation.  We provide some new examples of integrable systems with a vector potential whose existence can be discovered working in parabolic and elliptical coordinates. 

The plan of the paper is as 
follows: in section 2 we recall the structure of Hamiltonian systems 
with terms linear in the velocities; in section 3 we illustrate a 
version of the direct approach to find polynomial invariants which is 
particular efficient in treating two dimensional systems;  in section 
4, for sake of completeness, we recall systems admitting a second invariant which is a linear polynomial in the momenta;  in section 5 we treat the case of the 
quadratic second invariant invariant; in section 6 we illustrate all known strongly-integrable solutions in the quadratic case and in section 7 we conclude.
 
%%%%%%%%%%%%%%%%%%%%%%%%%%%%%%%%%%%%%%%%%%%%%%%%%%%%%%%%%%%%%%%%%%%%%%%%%%%%%%

\section{Hamiltonians with scalar and vector potentials}

We are interested in finding integrable examples of systems generated 
by a Hamiltonian function of the type 
\begin{equation}\label{Horig}
{\cal H} = \frac12 (p_x^2 + p_y^2) + A_{1} (x,y) \ p_x + A_{2} (x,y) \ p_y + V(x, y), 
\end{equation}
where the function $V$ is the ordinary ``scalar'' potential and $A_{1}$ and $A_{2}$ are the component of a ``vector'' potential ${\bf A}$ in two dimension. Under the canonical transformation 
\be 
p_x \longrightarrow p_x + \partial_{x} F, \;\; 
p_y \longrightarrow p_y + \partial_{y} F, \;\;
 \ee 
where $F(x,y)$ is an arbitrary function, the Hamiltonian remains form invariant if 
${\bf A}$ and $V$ are changed according to 
\ba 
{\bf A} &\longrightarrow& {\bf A} + {\bf \nabla} F, \\ 
V &\longrightarrow& V + {\bf A} \cdot {\bf \nabla} F +
 {\scriptstyle \frac12} |{\bf \nabla} F|^2. 
 \ea 
However, the two quantities 
\be\label{omega} \Omega(x,y) = 
{\scriptstyle \frac12} (\partial_{y}A_{1} - \partial_{x} A_{2}) \ee 
and 
\be\label{potef} W(x,y) = V - {\scriptstyle \frac12} |{\bf A}|^2 
\ee 
are ``gauge invariants" and can therefore be used to 
uniquely characterize the model system. The Hamiltonian to be worked 
on becomes then 
\be\label{H} {\cal H} = 
\frac12 (p_x+A_{1}(x,y))^2 + 
\frac12 (p_y+A_{2}(x,y))^2 + W(x,y). 
\ee  
$\Omega$, the ``curl'' of the vector potential, has several physical interpretations: in astrophysical and celestial mechanical applications, it usually denotes an angular velocity field; it is a magnetic field in electrodynamics and plasma physics and so on. We remark that in general it is easier to attempt to solve directly for $\Omega$ and $W$. To recover the scalar $V$, one must have $A_{1}$ and $A_{2}$ and this can be another difficult problem \cite{hiet}. 

We can first write the canonical equations provided by 
(\ref{Horig}),             
\ba 
\dot{x} &=& p_x + A_{1}, \\ 
\dot{y} &=& p_y + A_{2}, \\ 
\dot{p}_{x} &=& - \partial_{x} A_{1} \ p_x - \partial_{x} A_{2} \ p_y - \partial_{x} V, \\ 
\dot{p}_{y} &=& - \partial_{y} A_{1} \ p_x - \partial_{y} A_{2} \ p_y - \partial_{y}V,  
\ea 
and then simplify them by exploiting the functions introduced above 
to get the equations of motion 
\ba 
\ddot{x} -  2 \Omega \dot{y} &=& - \partial_{x} W, \label{motion1}\\ 
\ddot{y} + 2 \Omega \dot{x} &=& - \partial_{y} W. \label{motion2} 
\ea 
It is readily verified that under the 
phase-space flow generated by (\ref{motion1}--\ref{motion2}), there 
exists a conserved function that is the first invariant of the system {\it (Jacobi constant)}
\be\label{JJ} 
J = {\scriptstyle \frac12} (\dot{x} ^2 + \dot{y}^2) + W. 
\ee 

In the investigation 
of the integrability properties of Hamiltonian (\ref{H}), it turns 
out to be very helpful to work with complex variables. We perform 
then the canonical point transformation given by 
\ba
       z &=& x+iy,       \quad p_z = p = {\scriptstyle \frac12} (p_x - i p_y), \label{complex1}\\
 \bar z &=& x-iy,        \quad p_{\bar z} = \bar p = {\scriptstyle \frac12}  (p_x + i p_y), \label{complex2}
\ea
 so that Hamiltonian (\ref{H}) turns out to be 
 \be\label{Hcomplex} 
{\cal H} = 2 (p+\Phi) (\bar p+\bar \Phi) + W(z, \bar z), 
\ee 
where 
the complex function 
\be \Phi = {\scriptstyle \frac12} (A_{1} - i A_{2}) \ee 
has been introduced. In these variables, $\Omega$ is given by 
\be \Omega = 2 \Im\{\partial_{\bar z}\Phi\}, 
\ee 
where $\Im$ denotes the imaginary part. Equations 
(\ref{complex1}--\ref{complex2}) display a nice space-saving 
feature of using complex variables: even if an expression is not 
real, it is enough to write a single relation between complex 
functions (like, e.g. (\ref{complex1})). The remaining information 
is provided by the corresponding complex conjugate expression, which 
we therefore do not write explicitly. For example, the canonical 
equations given by (\ref{Hcomplex}) are 
\begin{eqnarray}
\dot{z} &=& 2  (\bar p+\bar \Phi), \label{ccan1}\\ 
 \dot{p} &=&-2 
(\bar p+\bar \Phi) \partial_{z} \Phi -2 (p+ \Phi) \partial_{z} {\bar \Phi}- \partial_{z} W, \label{ccan2}
\end{eqnarray}
 and the equations of motion corresponding to 
(\ref{motion1}--\ref{motion2}) are 
\be\label{cmotion} 
 \ddot{z} + 2 {i} \Omega \dot z = - 2 \partial_{\bar z} W.  
 \ee 
The Jacobi constant now is 
\be\label{cJ} 
 J = {\scriptstyle \frac12} \dot z \dot{\bar z} + W. 
 \ee  
%%%%%%%%%%%%%%%%%%%%%%%%%%%%%%%%%%%%%%%%%%%%%%%%%%%%%%%%%%%%%%%%%%%%%%%%%%%%%%

\section{Polynomial invariants}

We are working with a Hamiltonian system with two degrees of freedom 
of which we already know an invariant, the Jacobi constant. In order 
to identify integrable systems in the usual Liouville-Arnold sense, 
we have to find a second independent phase-space function conserved 
along the flow. The standard direct method to solve the problem 
consists in making a suitable ansatz about this function and trying 
to solve the system of differential equations ensuing by the 
conservation condition. For several reasons, the ansatz of a 
polynomial in the generalized momenta is the most common \cite{hiet}. It is well suited from the mathematical point of view, since 
it allows to get a system of PDEs in the coordinates only and is also 
well grounded on the basis of experience with already known 
integrable systems.  

Since we are looking for a real function, we 
make the following assumption 
\be\label{Minv} I^{(M)} = \sum_{k=0}^M 
(D_k (p+\Phi)^k + {\bar D}_k (\bar p+\bar \Phi)^k ). 
\ee 
Although it is common, in the vector potential case, to see the invariant written in terms of the velocities, this is the correct 
interpretation of $I^{(M)}$ as a phase-space function of the 
canonical variables.  
In order to satisfy the conservation of function (\ref{Minv}), we impose that its Poisson bracket with the Hamiltonian vanishes,
\be\label{conserv} 
\{ I^{(M)}, {\cal H}\} = 0 
\ee 
and try to solve for the 
complex functions $D_k(z, \bar z)$. In the presentation of the results, for easy comparisons with existing works, we will revert to the usual expressions in terms of the velocities, replacing the momenta in (\ref{Minv}) according to (\ref{ccan1}). In this case, conservation of the invariant can be checked by means of the condition
\be\label{iconserv} 
\frac{d I^{(M)}}{dt} = 0, 
\ee 
along the solution 
of the equations of motion (\ref{cmotion}).

Following the approach already 
used in the scalar case \cite{geom,max,kprs}, we consistently apply the trick of the energy constraint 
even in the present case. Here, with `energy', we mean the Jacobi constant (\ref{cJ}). This method has the advantage of allowing 
the simultaneous treatment of ``strong" invariants (the usual ones 
which are conserved for arbitrary values of the energy) and ``weak" 
invariants (functions which are conserved only on some energy 
surfaces). In the papers 
cited above, we have moreover shown how the energy constraint simplify 
the structure of the system of PDEs that has to be solved. The 
essential remark is that, to identify the cases of strong 
integrability, it is sufficient that in the final results a subset 
can be isolated which is independent of the energy parameter, in the 
present situation the given value of the first invariant, let us say 
$C$. If we are interested in a ``strongly-integrable system'', in the 
end we must get a solution independent of $C$. The procedure may 
appear more involute, but, at least in the scalar case, it reveals 
to be very effective.  

Formally, the idea of operating with the 
energy constraint is very simple. It consists first in introducing 
the ``null" Hamiltonian  
\be\label{nullH} 
{\cal H}_0 = {\cal H} - C 
\equiv 0, \ee 
or, on the same footing, the ``null" Jacobi invariant 
\be\label{nullJ} 
J_0 = J - C =\frac12 \dot z \dot{\bar z} - G \equiv 
0,  
\ee 
where $C$ as usual denotes the given fixed 
value of the Jacobi constant and
\be\label{JG} 
 G = C - W 
\ee 
is the so called ``Jacobi potential''. Second, using in a consistent way the constraint 
 \be\label{constraint} 
 2 (p+\Phi) (\bar p+\bar \Phi)  = G, 
 \ee 
wherever it appears in the computations. This essentially occurs 
when, implementing the conservation condition (\ref{conserv}), a 
polynomial relation in the generalised momenta appears and the constraint 
(\ref{constraint}) is used to eliminate powers of the mixed variables 
$(p+\Phi) (\bar p+\bar \Phi) $ in favour of powers of $G$. One minor shortcoming of this approach is that, in view of the explicit appearance of the function $G$, rather than simply the potential $W$, some of the coefficients of the invariant in general depend on the `energy' parameter too:
 \be\label{DC}
 D_{j} = D_{j} (z, \bar z; C), \quad 0 \le j \le M-2.
 \ee
 Therefore, to obtain the standard expression in terms of phase-space coordinates only, in the end we have to remember to perform the substitution 
 \be
 C \longrightarrow  
{\cal H} (p, \bar p, z, \bar z), 
\ee
wherever the parameter $C$ appears. In view of (\ref{DC}), we see that this replacement does not affect the degree of the polynomial in the momenta.
 
In practice, computing the Poisson bracket (\ref{conserv}), using the 
constraint, collecting the coefficients of the various powers of 
$p+\Phi$ (they are accompanied by their complex conjugates) and 
imposing their vanishing, we get the system 
\be\label{sistem} 
\partial_{\bar z} D_{k-1} + i k \Omega D_k + \frac{1}{2 G^k} 
\partial_z (G^{k+1} D_{k+1}) = 0, \quad k =0,1,...,M, \ee 
where it is 
implicitly assumed that $D_j = 0$ for $j<0$ and for $j>M$. The set of 
equation (\ref{sistem}) must be supplemented by the closure equations 
\be\label{cm} 
\partial_{\bar z} D_M = 0 \ee 
and 
\be\label{imm} 
\Re\{\partial_z (G D_1)\} = 0, \ee 
where $\Re$ denotes the real part. 
For sake of space, we do not write the expressions of the standard 
direct approach in real coordinates and without the energy 
constraint. To compare with, we refer to Section 4 of Hietarinta 
\cite{hiet} and recall the work of Hall \cite{hall}, where the study of weak invariants was first addressed and of Sarlet et al. \cite{SLC85a}, where some wrong deductions contained in Hall's work were corrected. A systematic analysis of the cases with $M=1$ (linear 
invariant) and $M=2$ (quadratic invariant) was started in Ref.\cite{dgrw} and recently taken up again in Ref.\cite{msw}. A more general version of the problem concerned with quadratic invariants is mentioned in a different context in Ref.\cite{FF}.

The first result one easily get with this approach is that 
equation (\ref{cm}) is readily solved as 
\be\label{SM} 
D_M = D_M (z), \ee 
that is, the leading order coefficient in the invariant is 
an {\it arbitrary analytical function}. This result, already known in 
the inertial case, still holds here. In Ref.\cite{geom}  
it was shown how, in the purely scalar case, the strong conservation condition restricts the 
form of this function. In what follows we will get analogous 
results when also a vector potential is present.   

One fundamental difference with what happens in the 
scalar case, is that equations for coefficients with even and odd 
indexes do not decouple. This fact is due to the Hamiltonian not be 
reversible in the present instance. System (\ref{sistem}) is 
therefore very awkward to solve in the general case. In the next two
sections we present the solutions in the linear and quadratic cases at fixed and arbitrary 
values of the Jacobi constant.  

%%%%%%%%%%%%%%%%%%%%%%%%%%%%%%%%%%%%%%%%%%%%%%%%%%%%%%%%%%%%%%%%%%%%%%%%%%%%%%

\section{Linear invariants}
  
We start the investigation looking for systems admitting a second 
invariant which is a linear function in the momenta. The ansatz is 
\be
\label{line}  I^{(1)} =  S (p+\Phi) + {\bar S} (\bar p+\bar \Phi) + K,   
\ee  
where for the three coefficients we have used a notation which 
conforms with that in previous works. The system of equations ensuing 
from the conservation condition is the following:  
\ba  
S_{\bar z} &=& 0, \label{c11}\\  
K_{\bar z} + i \Omega S &=& 0, \label{c12}\\ 
\Re\{(G S)_z\} &=& 0. \label{c13} 
\ea  
In order to compactify formulas,  from hereinafter with the subscript 
we denote the partial derivative with respect to the corresponding 
variable.
 
\subsection{The general solution for linear invariants}  

Equation (\ref{c11}) agrees with 
(\ref{cm}), confirming that $S$ can be an arbitrary analytic 
function, 
\be\label{SS} 
S = S (z). \ee 
To complete the treatment, 
given an arbitrary $S(z)$, we have to solve (\ref{c12}) and 
(\ref{c13}). This task is more efficiently achieved by performing a 
coordinate transformation that trivialize the differential equations. 
Let us consider a conformal transformation $z = F(w)$ to the new 
complex variable $w = X + i Y$ given by 
\be\label{cF1} \frac{dz}{dw} 
= F'(w) \equiv S(z(w)). \ee 
The explicit form of the transformation 
is then  
\be\label{cFi1} w = \int \frac{dz}{S(z)} \ee 
and we have 
the relation between the differential operators 
\be\label{cD1} 
\frac{d}{dw} = F'(w) \frac{d}{dz} = S \frac{d}{dz}. \ee 
Multiplying 
(\ref{c13}) by the real factor $S \bar S$, the content of the curly 
brackets can be modified in the following way: 
\be  S \bar S (S G)_z 
= S (S \bar S G)_z = (S \bar S G)_w. \ee 
Introducing the ``conformal'' 
potential  
\be\label{cpot} \widetilde G = |F'|^2 G = |S|^2 G = S \bar S 
G, \ee 
eq.(\ref{c13}) reduces to 
\be \Re\{{\widetilde G}_w\}=0, \ee 
which is readily solved in 
\be\label{gs} \widetilde G = g (Y), \ee 
where 
$g$ is an arbitrary real function and, according to the definition of 
the coordinate transformation, $Y$ is the imaginary part of $w$.   

In 
the new variables $w$, eq.(\ref{c12}), together with its complex 
conjugate, can be rewritten as 
\ba 
K_{w}        &=&   i \widetilde \Omega, \\ 
K_{\bar w} &=& - i \widetilde \Omega, 
\ea 
where the conformal field 
\be
 \widetilde \Omega = |S|^2 \Omega  
 \ee
  has been introduced. 
The integrability condition 
for the real function $K$ is 
\be \Re\{K_w\}=0, \ee 
with solution 
\be\label{ks} K = k (Y), \ee 
where $k$ is another arbitrary real 
function. The conformal vector potential is then given by 
\be\label{os} 
\widetilde \Omega = - \frac{k'(Y)}{2}. \ee 
An inversion of the coordinate 
transformation allows to express the solution in the original 
variables. The second invariant can be expressed as 
\be\label{I1} 
I^{(1)} = \Re\{S\} \dot x + \Im\{S\} \dot y + K. \ee  

\subsection{Linear invariants at arbitrary energy}

Equation (\ref{c13}), in view of (\ref{SS}) and recalling the 
definition of $G$ in (\ref{JG}), can be rewritten as 
\be\label{G1} \Re\{S'(C-W) - S W_z\}=0. \ee 
If we are interested in 
strong integrability, namely in an invariant which is conserved for 
arbitrary values of the Jacobi constant, equation (\ref{G1}) must be 
independent of $C$. Therefore, it decouples in two independent 
equations: the first is 
\be\label{S1} \Re\{S'(z)\}=0. \ee  
The 
second turns out to be 
\be\label{W1} \Re\{(S W)_z\}=0. \ee 
Eq.(\ref{S1}) means that at arbitrary energy, we are no longer free in the choice of the 
coordinate transformation: we have to comply 
with this condition which actually imposes very strong limitations. 
It can be integrated to give 
\be\label{Sexp1} S(z)=i k z + \alpha, 
\ee  
where $k$ is a real constant and $\alpha$ a complex constant. We 
can prove that it essentially allows only two kind of new 
coordinates: a) polar coordinates; b) rotated Cartesian coordinates. 
To show this, we first observe that we can exploit translations and 
scaling of the complex plane to further reduce the freedom contained 
in (\ref{Sexp1}). If $k$ is not zero, a translation allows to put 
$\alpha=0$. A scaling allows then to put $k=1$. We have then the two 
possibilities: 
\ba {\rm a}) \quad S(z) &=& iz, \quad F(w) = {\rm 
e}^{i w}, \quad x = e^{-Y} \cos X, \quad y= e^{-Y} \sin X,\\ {\rm b}) 
\quad S(z) &=& \alpha, \quad F(w) =\alpha w, \quad x = a X - b Y, 
\quad y=  b X + a Y, \ea 
Case a) can be recognized as the 
transformation to {\it polar} coordinates. In fact, with the usual 
notation, they are defined as 
\be\label{polar} 
r =  {\rm e}^{-Y}, 
\quad \theta = X. 
\ee 
Solutions (\ref{gs}--\ref{os}), in view of 
(\ref{nullJ}), are 
\be\label{solpolar1} 
W = W(r) = \frac{g(r)}{r^2}, \quad K = 
k(r), \quad \Omega = \frac{k'(r)}{2 r}
\ee 
and the second invariant 
turns out to be 
\be \label{solpolar2}
I^{(1)} = i (z \dot z - {\bar z} \dot{\bar z}) + 
K = r^{2} {\dot \theta} + k(r). 
\ee 
The problem is rotationally 
symmetric and the corresponding invariant is a generalisation of the angular momentum.  

Case b) can be recognized as the transformation to {\it rotated 
Cartesian} coordinates. Solutions (\ref{gs}--\ref{os}), in view 
of (\ref{nullJ}), then gives  
\be W = W(a y - b x), 
\quad K = k(a y - b x), 
\quad \Omega = -\frac{k'(a y - b x)}{2}. 
\ee 
The second 
invariant now is  
\be 
I^{(1)} = \dot z + \dot{\bar z} + K = 
a \dot x + b \dot y + k(a y - b x). 
\ee 
The problem is invariant under 
translation along the family of straight lines $a x + b y = const.$ 
These two solutions are already well known \cite{dgrw}, 
and the above procedure can be appreciated in its effectiveness.    

\subsection{Examples of weakly integrable systems with linear 
invariants}  

We may provide two interesting classes of weakly 
integrable systems admitting linear invariants.   

The first is 
obtained by the simple observation that, if we chose the level 
surface $C=0$, it is no more necessary that condition (\ref{S1}) be 
satisfied. {\it Any} analytic function $S=S(z)$ provides a solution 
through the corresponding conformal transformation. If $Y$, as above, 
denotes the new coordinate 
\be Y = \Im\left\{\int 
\frac{dz}{S(z)}\right\}, \ee 
then the solution is given by 
\be W = 
\frac{g(Y(x,y))}{|S|^2}, \quad K = k(Y(x,y)), \quad \Omega = 
-\frac{k'(Y(x,y))}{2 |S|^2}, \ee 
with $g$ and $k$ arbitrary real 
functions. 

The second 
class of weakly integrable systems is obtained with the following 
trick. Let us consider the analytic function $f(z)$ and consider then 
the conformal transformation (\ref{cF1}) with $S(z)$ given by 
\be\label{sweak} 
S = \frac{1}{c + f'(z)}, \ee 
with $c$ constant. 
Recalling definition (\ref{cpot}), let us consider the ``flat" 
conformal potential $\widetilde G = 1$. In this case, relation 
(\ref{cpot}), using (\ref{sweak}), gives 
\be G = C - W = 
\frac{1}{|S|^2} = c^2 + c (f'+ \bar f') + |f'|^2. \ee 
We can 
therefore interpret $c^2$ as the fixed value of the Jacobi constant 
\be
c^2 \equiv C
\ee
and get as a consequence a {\it C-dependent} potential  
\be 
W(z, \bar z; C) = - \sqrt{C} (f'+ \bar f') - |f'|^2.  
\ee 
To complete the 
solution, we have to write explicitly the coordinate transformation 
generated by (\ref{sweak}), that is 
\be 
w = \int \frac{dz}{S(z)} = c 
z + f(z). \ee 
Again, an arbitrary function $k(Y)$, with 
\be Y = 
\Im\left\{c z + f(z)\right\}, \ee 
will do the work. For further details on these systems we refer to Ref.\cite{gp}.

%%%%%%%%%%%%%%%%%%%%%%%%%%%%%%%%%%%%%%%%%%%%%%%%%%%%%%%%%%%%%%%%%%%%%%%%%%%%%%

\section{Quadratic invariants}
  
We now look for systems admitting a second invariant which is a 
quadratic function in the momenta. The ansatz is 
\be\label{quad}  
I^{(2)} = S (p+\Phi)^{2} + {\bar S} (\bar p+\bar \Phi)^2 + 
            R (p+\Phi) + \bar R (\bar p+\bar \Phi) + K,  
            \ee  
where, besides $S$ and the real function $K$, 
we now have to determine the complex function $R$. The system of 
equations ensuing from the conservation condition is the following:  
\ba  
S_{\bar z} &=& 0, \label{c21}\\  
R_{\bar z} + 2 i \Omega S &=& 0, \label{c22}\\  
K_{\bar z} + S G_z + {\scriptstyle \frac12} S' G + i \Omega R &=& 0, \label{c23}\\ 
\Re\{(R G)_z\} &=& 0. \label{c24} 
\ea 
where, in (\ref{c23}) we have already exploited (\ref{c21}), 
that, as usual, embodies the fact that $S$ is an arbitrary analytic function.   

\subsection{Towards a general solution for quadratic invariants}  

System (\ref{c21}--\ref{c24}) is much more difficult to solve than 
the previous linear case. Indeed we lack of a general solution. The main 
reason for this difficulty is that the coupling between $G$, $R$ and 
$\Omega$ produces an integrability condition for $K$, through 
eq.(\ref{c23}) and its complex conjugate, that is a {\it nonlinear} 
PDE. However, we can implement the strategy to arrive as close as possible to a general solution and, what is of great importance, we can solve the problem in the strongly integrable case, developing an effective way to construct solutions.
           
           This time we 
use a conformal transformation to the complex variable $w = X + i Y$ 
given by 
\be\label{cF2} 
\frac{dz}{dw} = F'(w) \equiv \sqrt{S(z(w))} 
\ee 
so that the explicit form of the transformation is 
\be\label{cFi2} 		
w = \int \frac{dz}{\sqrt{S(z)}}. 
\ee 
Introducing 
the conformal potential  
\be\label{cpot2} 
 \widetilde G = |F'|^2 G = |S| G = \sqrt{S \bar S} G, 
 \ee
and the new complex function
\be\label{TR}
  \widetilde R = \frac{R}{\sqrt{S}},
  \ee
  eq.(\ref{c24}) keeps its form in the transformed coordinates
  \be\label{T24} 
  \Re\{(\widetilde R \widetilde G)_w\}=0. 
  \ee
 The solution of this equation is
 \be\label{solT24}
 \widetilde R \widetilde G = i {\cal K}_{\bar w},
 \ee
 where $\cal K$ is an arbitrary real function.
 
 Let us now multiply both sides of (\ref{c22}) by
 \be
  \sqrt{\frac{\bar S}{S}}.
  \ee
 Using (\ref{TR}) and introducing the conformal field 
 \be\label{comega2}
 \widetilde \Omega = |S| \Omega,
 \ee
eq.(\ref{c22})  transforms into
\be\label{T22} 
\widetilde R_{\bar w} + 2 i \widetilde \Omega = 0.
\ee
Since the conformal field is real, the solution of (\ref{T22}) is
\be\label{solT22}
 \widetilde R = - 4 i \xi_{w},
 \ee
where $\xi$ is another arbitrary real function. The factor $-4$ appears for later convenience. In this way, the conformal field is given by
\be\label{solTomega2}
 \widetilde \Omega = 2 \xi_{w \bar w}.
 \ee

At this point, to get the general solution of the unified treatment of weak and strong integrability, one should try to solve the integrability condition for eq.(\ref{c23}) in the light of the results (\ref{solT24}--\ref{solT22}). Let us write the  
integrability condition for $K$ which, computing $K_{z \bar z} = 
K_{\bar z z}$ from (\ref{c23}), is 
\be\label{ntcond} 
\Im\{S'' G + 3 
S' G_z + 2 S G_{zz} + 2 i (\Omega R)_z\} =0. 
\ee 
In the transformed coordinates, this becomes
\be\label{ntcond2} 
\Im\{\widetilde G_{ww}\} + \Re\{(\widetilde R \widetilde \Omega)_w\} =0. 
\ee 
>From (\ref{solT24}) and (\ref{T22}) we have 
\begin{equation}
   \widetilde R \widetilde\Omega = \textstyle\frac{i}2 \widetilde R \widetilde R_{\bar w}
    = \textstyle\frac{i}4 (\widetilde R^2)_{\bar w}
    = -\textstyle\frac{i}4 (\widetilde G^{-2} {\cal K}_{\bar w}{}^2)_{\bar w}
\end{equation}
so that (\ref{ntcond2}) can be rewritten as
\begin{equation}
   \Im\{\widetilde G_{ww}\}
    = \textstyle\frac14 \Re\{ i(\widetilde G^{-2}{\cal K}_{\bar w})_{\bar w w} \}
\end{equation}
We can try to solve this equation after specifying $\cal K$. Unfortunately, the equation is highly nonlinear except in the rather trivial case when $\cal K$ is constant, so it appears to be very difficult to solve it in the general case. We direct our attention attempting to solve the more limited but fundamental case of strong integrability.

\subsection{Quadratic invariants at arbitrary energy} 

Starting again with (\ref{c24}), we see that, in view of definition (\ref{JG}), it can be 
rewritten as 
\be\label{G2} \Re\{R_z(C-W) - R W_z\}=0. \ee 
If $R$ is 
independent of $C$, as is the case under study, in order to satisfy this equation at 
arbitrary values of the energy, it decouples in two independent 
equations: the first is 
\be\label{R2} \Re\{R_z\}=0; \ee  
the second
\be\label{W2} \Re\{R W_z\}=0. \ee 
Eq.(\ref{R2}) is 
solved by introducing the arbitrary real function $\eta$ so that 
\be\label{RK} 
R = - 4 i {\eta}_{\bar z}. \ee 
This allows to solve 
equation  (\ref{W2}) for $W$ in the form 
\be\label{Weta} W=W({\eta}), \ee 
that is $W$ is, at this stage, an arbitrary function of the argument.

Comparing (\ref{RK}) with (\ref{solT22}) and taking into account definition 
(\ref{TR}), we get the relation
\be\label{xieta1}
\sqrt{S \bar S} \xi_{w} = |S| \xi_{w} = \eta_{\bar w}. \ee
Solving the integrability condition for $\eta$ with a given form of $S$ allows to find $\xi$ (and viceversa): equation (\ref{xieta1}) is deceivingly simple; we will see later that it is of some concern. Here we remark that the route followed to treat eq.(\ref{c24}) reverted to the original physical coordinates $z, \bar z$, since they lead to the simple result in (\ref{RK}). However, eq.(\ref{solTomega2}) above is valid in general: it is the simplest way in which we can solve for the vector potential and is expressed in the new coordinates $w, \bar w$. This result and the development below show how working in the new coordinates is advantageous in this context too.

In the meantime we have to determine the forms of $S$ imposed by strong integrability. Examining the integrability condition (\ref{ntcond}), the usual condition of independence of the results 
from $C$ imposes the following constraints on the form of $S$ 
\be\label{S2} \Im\{S''(z)\}=0. \ee  
This result, which is valid in 
the scalar case also, is a natural extension of what found in the 
linear case.    

Coming back to eq.(\ref{xieta1}), we note that, using real coordinates, we have
\ba
|S| \ \xi_{X} &=& \eta_{X},\label{xieta21}\\
|S| \ \xi_{Y} &=& -\eta_{Y}.\label{xieta22}
\ea
It can be proven \cite{pr} that condition (\ref{S2}) is equivalent to
\be\label{SXY}
|S|_{XY}=0,
\ee
so that we can also write
\be\label{sepS}
|S| = A(X) + B(Y),
\ee
with $A$ and $B$ determined by the specific form of $S$. Therefore, eqs.(\ref{xieta21}--\ref{xieta22}) generates the following differential equations for the functions $\eta$ and $\xi$
\ba
A' \eta_{Y} + B' \eta_{X} - 2 (A + B) \eta_{XY} &=& 0, \label{xe1}\\  
A' \xi_{Y} + B' \xi_{X} + 2 (A + B) \xi_{XY} &=& 0. \label{xe2}
\ea
In 
view of  (\ref{RK}), that can be rewritten as
\be\label{RK2} 
|S| \widetilde R = - 4 i {\eta}_{\bar w} \ee 
and of (\ref{comega2}), integrability condition (\ref{ntcond2}) can be written in the form  
\be 
\Im\{\widetilde G_{ww}\} + |S| \ \Re\{\widetilde R \Omega_w\} =0,
\ee 
or, in real coordinates,
\be\label{ntcond3} 
\widetilde G_{XY} + 2 (\eta_{X} \Omega_Y - \eta_{Y} \Omega_X)=0. 
\ee 
If we want to exploit the result established by (\ref{Weta}), we have to resume the physical potential $W$ through
\be
\widetilde G = |S| (C - W).
\ee 
Using (\ref{sepS}) we have
\be
\widetilde G_{XY} = - A' W_{Y} - B' W_{X} - (A+B) W_{XY} 
\ee
and so (\ref{ntcond3}) is
\be\label{ntcond4} 
(A+B) W_{XY} + A' W_{Y} + B' W_{X} = 2 (\eta_{X} \Omega_Y - \eta_{Y} \Omega_X).
\ee
But (\ref{Weta}) tells us that
\be
W_{X} = \eta_{X} W'(\eta) , \quad
W_{Y} = \eta_{Y} W'(\eta) , \quad 
W_{XY} = \eta_{X} \eta_{Y} W'' + \eta_{X Y} W' , \ee
so that (\ref{ntcond4}) becomes
\be\label{ntcond5} 
W''(\eta) + 3 \frac{\eta_{X Y}}{\eta_{X} \eta_{Y} }W'(\eta) = \frac{2}{A+B} 
\left( \frac{\Omega_Y}{\eta_{Y}} - \frac{\Omega_X}{\eta_{X}} \right).
\ee 
This is the best form we attain to express the integrability condition for the potential: we see that it implies that 
\be\label{condeta}
\frac{\eta_{X Y}}{\eta_{X} \eta_{Y} } = \Phi (\eta),
\ee
where $\Phi$ is arbitrary. 

Therefore, the strategy to find strongly integrable systems with scalar and vector potentials supporting quadratic invariants is: to chose a suitable $S$ in the class determined by condition (\ref{S2}) (and therefore $A$ and $B$) and solve (\ref{xe1}--\ref{xe2}) to find $\eta(X,Y)$ and $\xi(X,Y)$; to solve (\ref{solTomega2}) to find \be\label{solTomegaXY}
\widetilde \Omega = \frac12 (\xi_{XX} + \xi_{YY})
\ee 
and use (\ref{comega2}) to find 
\be\label{comega2XY}
\Omega (X,Y) = \frac{\widetilde \Omega}{A+B}
\ee 
and, finally, try to solve (\ref{ntcond5}) for 
\be\label{ntcond5XY} 
W = W\left(\eta(X,Y)\right)
\ee 
taking into account (\ref{condeta}).

\subsection{General form of the quadratic invariant} 

In the next section we apply the strategy delineated above illustrating how already known and new integrable systems are determined. We end this section with a closer look at the structure of the quadratic invariant.

In the new variables, $w, \bar w$, together with the conformal transformation, it is natural to introduce the new time variable $\tau$ such that
\be\label{newtime}
d\tau = {{dt} \over {|F'|^{2}}}. 
\ee
We can use the apex to also denote the derivative with respect to $\tau$ without risk of confusion with other derivatives with respect to  coordinates. Equations of motion (\ref{cmotion}) then assume the form:
\be\label{cwmotion} 
 w'' + 2 {i} \widetilde \Omega w' = 2 \widetilde G_{\bar w}.  
 \ee 
In view of above positions, the quadratic invariant is most simply expressed as
\be\label{quad2C}  
I^{(2)} = \Re \left\{ \frac12 \left(w' \right)^2 + 
                                 {\widetilde R} {\bar w}' \right\} + K,  
            \ee  
or, in terms of real variables,
 \be\label{quad2R}  
I^{(2)} = \frac12 \left[ \left(X' \right)^2 - 
                          \left(Y' \right)^2 \right] 
                   - 2  (\xi_{X} Y' + \xi_{Y} X') + K.  
            \ee  
We recall that the scalar $K$, to be found by integrating the transformed version of (\ref{c23}), namely
\be
K_{\bar w} + \widetilde G_w + i \widetilde \Omega {\widetilde R}  = 0, 
\ee
is a function of the form
\be
K = K(w, \bar w; C).
\ee
In the present case of strong integrability, in view of (\ref{sepS}), it turns out that this function can be expressed as
\be\label{KK}
K = k + C(B-A),
\ee
and $k$ can be found by integrating the system
\ba
k_{X} - \widetilde W_{X} + 4 \widetilde \Omega \xi_{X} &=& 0,\label{k1}\\
k_{Y} + \widetilde W_{Y} - 4 \widetilde \Omega \xi_{Y} &=& 0, \label{k2}
\ea
 where $\widetilde W = (A+B)W$. From (\ref{quad2R}), using (\ref{KK}) and
 \be
 C = \frac12 \frac{\left(X' \right)^2 + \left(Y' \right)^2}{A+B} + W,
 \ee
 the general form of the invariant is then
 \be\label{quad2RT}  
I^{(2)} = \frac{1}{A+B} 
              \left[ B \left(X' \right)^2 - A 
              \left(Y' \right)^2 \right] 
             - 2  ( \xi_{X} Y' + \xi_{Y} X' ) + (B-A)W + k.  
            \ee
Observing that the relations between velocities in the two gauges are given by
\ba
X'  &=& \Re \{F'\} {\dot x} + \Im \{F'\} {\dot y}, \label{veltr1}\\
Y'  &=& - \Im \{F'\} {\dot x} + \Re \{F'\} {\dot y},\label{veltr2}
\ea
we can eventually transform the invariant in Cartesian coordinates. Transformation rules automatically account also for the change of the time variable according to (\ref{newtime}).

%%%%%%%%%%%%%%%%%%%%%%%%%%%%%%%%%%%%%%%%%%%%%%%%%%%%%%%%%%%%%%%%%%%%%%%%%%%%%%

\section{Solutions with quadratic invariants}

Let us recall the 
coordinate systems given by the condition for the existence of strong quadratic invariants.  We observe that the function $S$ we have to use 
is obtained by integrating (\ref{S2}) so that 
\be\label{Sexp2} 
S(z)=c z^2 + \beta z + \alpha, \ee  
where $c$ is a real constant and 
$\beta, \alpha $ complex constants. Exploiting the freedom of making 
translations, rotations and scaling in the complex plane, we have the 
following four inequivalent cases (for further details we refer to Refs.\cite{geom,SPT}): 
\begin{eqnarray}
{\rm a}) \quad S(z) &=& \alpha, \quad F(w) =\alpha w, \quad x = a X - 
b Y, \quad y=  b X + a Y,\label{Scases1}\\ 
{\rm b}) \quad S(z) &=& z^2, \quad F(w) = 
e^{w} , \quad x = e^{X} \cos Y, \quad y= e^{X} \sin Y,\label{Scases2}\\ 
{\rm c}) 
\quad S(z) &=& 4z, \quad F(w) = w^2, \quad x = X^2 - Y^2, \quad y= 
2XY,\label{Scases3}\\ 
{\rm d}) \quad S(z) &=& z^2 + \Delta^2, \quad F(w) = \Delta 
\sinh{w} ,              \quad x =  \Delta \sinh {X} \cos Y, 
\quad y=  \Delta \cosh {X}  \sin Y. \label{Scases4}
\end{eqnarray}
  
Case a) gives the rotated Cartesian coordinates (standard Cartesian 
coordinates if $\alpha = 1$). Case b) gives again the polar 
coordinates. Case c) gives the parabolic coordinates which are also 
referred to as {\it Levi-C\`{\i}vita} coordinates (the factor 4 in 
the definition of $S$ appears just for this reason). Finally, case d) 
produces the elliptical coordinates. For sake of completeness, we list, for each of the four cases, the conformal factor and functions $A$ and $B$ introduced in (\ref{sepS}):
\ba
{\rm a}) \quad |S| &=& 1, \quad A = B = \scriptstyle{\frac12},\label{SAB21}\\
{\rm b}) \quad |S| &=& e^{2X}, \quad A = e^{2X}, \quad B = 0,\label{SAB22}\\
{\rm c}) \quad |S| &=& 4 (X^{2} + Y^{2}), \quad A = 4 X^{2}, \quad B = 4 Y^{2},\label{SAB23}\\
{\rm d}) \quad |S| &=& \Delta^{2} (\sinh^{2} {X} + \cos^{2} Y), 
\quad A = \Delta^{2} \sinh^{2} {X} , 
\quad B = \Delta^{2} \cos^{2} Y.\label{SAB24}
\ea

\subsection{Cartesian coordinates}

Let us start with the 
simplest case a), the Cartesian one. As the structure of the solution 
will show in the end, there is no need to work with rotated 
coordinates, therefore we can put $\alpha = 1$ and make the trivial identification $X=x$ and $Y=y$. 
Eqs.(\ref{xieta21}--\ref{xieta22}) are
\ba
\eta_{x} &=& \xi_{x} , \label{xe11}\\  
\eta_{y} &=& - \xi_{y} . \label{xe12}
\ea
The solution is
\ba
\xi &=& f(x) + g(y),\\
\eta &=& f(x) - g(y),
\ea
with $f$ and $g$ arbitrary. From (\ref{solTomegaXY}) and (\ref{comega2XY}), the vector field is
\be\label{omega21}
\Omega = \frac12 (f'' + g'')
\ee
and the $R$ function
\be\label{R21}
R = - 2 (g' + i f').
\ee
Eq.(\ref{ntcond4}) is
\be\label{ntcond41} 
W_{xy} - (g' f''' + f' g''')=0.
\ee
>From (\ref{Weta}) we have that $W = W(f(x) - g(y))$, so that
\be
W_{xy} = - f' g' W'' 
\ee
and (\ref{ntcond41}) becomes
\be
W'' + \frac{f'''}{f'} + \frac{g'''}{g'} = 0.
\ee
A further differentiation by $x$ and $y$ gives
\be
\frac{1}{f'} \frac{d}{dx} \frac{f'''}{f'} = - \frac{1}{g'} \frac{d}{dy} \frac{g'''}{g'} =
2 a = -W'''
\ee
with $a$ real constant. We then get two equations for $f$ and $g$
\ba
f'' &=& a f^{2} + b f + c,\\
g'' &=& - a g^{2} + d g + e,
\ea
in agreement with the work of Dorizzi et al. \cite{dgrw}, that can be integrated for various choices of the constants $a,b,c,d,e$. The potentials in terms of $f$ and $g$ are
\ba
W &=& \frac{a}{3} (g-f)^{3} - \frac{b+d}{2} (g-f)^{2} + (c-e+m) (g-f),\\
2 \Omega &=& a (f^{2} - g^{2}) + b f + d g + c + e,
\ea
where $m$ is another constant. Integrating eqs.(\ref{k1}--\ref{k2}) gives then
\be
K = a fg (f+g) + (b-d) fg - a (f^{3} + g^{3}) - \frac12 (3b+d) f^{2} + \frac12 (3d+b) g^{2} - (3c + m + e)f + (c - m +3e) g.
\ee
  
  The fundamental example of the harmonic oscillator is obtained with the choice
  \be
  a = b = d = 0,
  \ee
  so that
  \ba
f &=& \frac12 c x^{2},\\
g &=& \frac12 e y^{2}
\ea
and the potentials are
\ba
W &=& \frac12 (c-e+m)  (e y^{2}-c x^{2}),\\
2 \Omega &=& c + e.
\ea
We remark that, in this case, the vector field is a {\it constant angular velocity} and that the isotropic oscillator only exists when the angular velocity vanishes. The second invariant in this case is
 \be\label{quad2carT}  
I^{(2)} = \frac12 
              ( \dot x^2 - \dot y^2)
             - 2  ( c x \dot y + e y \dot x ) + \frac12 
             \left[ (c - m +3e) e y^{2} - (3c + m + e)c x^{2} \right].  
            \ee

\subsection{Polar coordinates}

In case b), the polar coordinates, we recall that now 
\be\label{polar2} 
r =  {\rm e}^{X}, \quad \theta = Y
\ee 
and remark the difference of this relation with respect to (\ref{polar}) obtained in the linear case. We have
\be
|S| = A =  {\rm e}^{2X}= r^{2}, \quad B = 0.
\ee
Eq.(\ref{xe1}) becomes
\be
2 {\rm e}^{2X} (\eta_{X Y} - \eta_{Y}) = 0,
\ee
which is solved by
\be
\eta =  {\rm e}^{X} g(Y) + f(X),
\ee
with $f$ and $g$ arbitrary. From (\ref{xieta1}) we then have
\be
\xi = - {\rm e}^{X} g + \int  {\rm e}^{-2X} f'(X) dX.
\ee
>From (\ref{solTomegaXY}) and (\ref{comega2XY}), the velocity field is
\be\label{omegapolar}
2 \Omega =  {\rm e}^{-2X} \Delta \xi =  
                     {\rm e}^{-4X} (f'' - 2 f') -  {\rm e}^{-3X} (g'' + g).
\ee
Condition (\ref{condeta}) says that
\be\label{condetapolar}
\eta_{X} =  {\rm e}^{X} g (Y) + f' (X) = \Phi({\rm e}^{X} g(Y) + f(X)).
\ee
There are two ways in which (\ref{condetapolar}) can be accomplished: the first is $g={\rm const}$ and it is easy to see that in this way we are actually taken back to the spherically symmetric case to which actually a linear invariant is associated. We can take $\eta(X)$ and $W(X)$ arbitrarily whereas, to find the vector field, we observe that
\be
\xi'(X) =  {\rm e}^{-2X} \eta'(X) 
\ee
so that 
\be\label{omegapolar2}
2 \Omega =  {\rm e}^{-2X} \xi'' =  
                     {\rm e}^{-4X} (\eta'' - 2 \eta') =
                     {\rm e}^{-2X} ({\rm e}^{-2X} \eta')'.
\ee
To show that this solution is equivalent to that obtained in the linear case, we compute the scalar $k$ by integrating (\ref{k1}) to get
\be
k^{(2)}(X) = {\rm e}^{2X} W - ({\rm e}^{-2X} \eta')^{2}.
\ee
Observing that, in polar coordinates, relation (\ref{newtime}) between old and new time is 
\be
d\tau = {{dt} \over {r^{2}}}
\ee
and using
\be
Y' = r^{2} {\dot \theta}, \ee 
expression (\ref{quad2RT}) becomes
 \be\label{quad2polar}  
I^{(2)} = - r^{4} {\dot \theta}^{2} - 2 \eta' {\dot \theta} - 
 \frac{(\eta')^{2}}{r^{4}}.
 \ee
In the linear case, recalling solution (\ref{solpolar2}), we got:
\be\label{solpolar3}
I^{(1)} =r^{2} {\dot \theta} + k^{(1)}(r), 
\ee
where, from (\ref{solpolar1}),
\be
k^{(1)}(r) = 2 \int r \Omega d r = 
                 2 \int {\rm e}^{2X} \Omega dX.
                 \ee
  Using (\ref{omegapolar2}) this becomes
  \be
  k^{(1)}(r) = {\rm e}^{-2X} \eta' = r^{-2} \eta' ,
  \ee
  so that, comparing (\ref{quad2polar}) with (\ref{solpolar3}), we see that
  \be
  I^{(2)} = - (    I^{(1)} )^{2}   .\ee

The second possibility of satisfying condition (\ref{condetapolar}) is to have $f={\rm const}$ and the constant, without any loss of generality, can be put equal to zero. Therefore, we have
\be\label{etapolar}
\eta =  {\rm e}^{X} g (Y) 
\ee
and
\be\label{xipolar}
\xi = - {\rm e}^{-X} g .
\ee
To find the potential, we have to solve (\ref{ntcond5}) which, using (\ref{etapolar}) and (\ref{omegapolar}), is
\be\label{ntcond5polar} 
W''(\eta) + \frac{3}{\eta} W'(\eta) = - \frac{1}{{\rm e}^{6X} g g'} 
\left( g''' g + 3 g'' g' + 4 g' g \right).
\ee 
The simplest way in which the right hand side too is a function of $\eta$ is given by the condition
\be\label{gpol}
g''' g + 3 g'' g' + 4 g' g  = \alpha \frac{g'}{g^{5}},
\ee
with $\alpha$ constant, so that (\ref{ntcond5polar})
becomes
\be\label{ntcond5polar2} 
W''(\eta) + \frac{3}{\eta} W'(\eta) + \frac{\alpha}{{\eta}^{6}} 
=0
\ee
with solution
\be\label{ntcond5polarsol} 
W = \frac{\beta}{{\eta}^{2}} - \frac{\alpha}{8 {\eta}^{4}}.
\ee
To find explicit solution we have to determine the possible forms of $g$ which can be got by integrating (\ref{gpol}) twice 
\be\label{gpol2}
(g')^{2} + g^{2} + \frac{\alpha}{4 g^{4}} + \frac{\gamma}{g^{2}} = \delta,
\ee
with $\gamma$ and $\delta$ constants. All this is in agreement with the results in McSween and Winternitz \cite{msw} to which we refer. 

Since in the approach with the energy constraint, the coefficients appearing in the invariant must be supplemented by a further substitution in order to get the final physical expression, for sake of completeness, we work out a specific example. To ease the comparison with the results in Ref.\cite{msw}, we use explicitly polar coordinates as in (\ref{polar2}).
The simplest case in the solution above is given by the choice $ \alpha = 0 $. Eq.(\ref{gpol2}) has solution
\be\label{gpol3} 
g(\theta) = \sqrt{a + b \cos 2 \theta} \ee 
with the constants $a$ and $b$ given by
\be
\gamma = a^{2} - b^{2}, \quad \delta = 2 a.
\ee
>From (\ref{ntcond5polarsol}) and (\ref{omegapolar}), the potential and velocity field are given respectively by
\be
W = \frac{\beta}{r^{2}(a + b \cos 2 \theta)}
\ee
and
\be\label{omepol3} 
\Omega =  \frac{b^{2} - a^{2}}{2 r^{3}(a + b \cos 2 \theta)^{3/2}}.
\ee
To find the invariant, we can use (\ref{quad2RT}). Eqs.(\ref{k1}--\ref{k2}) can easily be integrated to get
\be
k = - r^{2} W - r^{-2} g (g + g'').
\ee
so that
expression (\ref{quad2RT}) becomes
 \be\label{quad2polar1}  
I^{(2)} = r^{4} {\dot \theta}^{2} + 2 (r g {\dot \theta} - {\dot r} g') +
 \frac{2 \beta}{g^{2}} + \frac{g (g + g'')}{r^{2}}.
 \ee
 In the specific example of (\ref{gpol3}) we get
 \be\label{quad2polar2} 
I^{(2)} = r^{4} {\dot \theta}^{2} + 
 \frac{2 b \sin \theta}{\sqrt{a + b \cos 2 \theta}}{\dot r} + 
 2 \sqrt{a + b \cos 2 \theta} r {\dot \theta} + 
 \frac{1}{a + b \cos 2 \theta} \left(\frac{a^{2} - b^{2}}{r^{2}} + 2 \beta \right).
 \ee
We remark that the small discrepancy in (\ref{quad2polar2}) with respect to the analogous expression reported in Ref.\cite{msw} is due to a difference of a factor 2 in the definition of $\Omega$ as can be seen from eq.(\ref{omepol3}).

\subsection{Parabolic coordinates}

We use the Levi-C\`{\i}vita representation of the parabolic coordinates introduced in case c) above and mostly used in celestial mechanical applications. As a first step, the general strategy we depicted at the end of subsection 5.2 prescribes to determine the functions $\eta$ and $\xi$ satisfying equations (\ref{xe1}--\ref{xe2}). According to (\ref{SAB23}), we have
\be
A = 4 X^{2}, \quad B = 4 Y^{2}.
\ee
so that eq.(\ref{xe1}) becomes
\be\label{etap}
X \eta_{Y} + Y \eta_{X} - (X^{2} + Y^{2}) \eta_{XY} = 0.
\ee
A fairly general solution of this PDE  can be represented in the following form
\begin{equation}\label{etaparagen}
   \eta(X,Y) = \int F(a) \sqrt{(X^2+a)(Y^2-a)} d a +
                     \int G(a) \frac{(X^2+Y^2)^{2}}
                        {\left((X^2+a)(Y^2-a)\right)^{3/2}} d a,
\end{equation}
where $a$ is an arbitrary real parameter and $F$ and $G$ two arbitrary smooth functions. 

The subsequent steps should consist in determining the functions $\xi$ and $\Omega$ and finally to solve for the potential $W(\eta)$. However, we observe that it is not easy to satisfy 
the constraint (\ref{condeta}) with a too general expression for $\eta$: 
therefore, we first find a suitable form of this function and then 
proceed as above. A simple but non trivial possibility is that given by the position
\be
F = c \delta(a-b), \quad G = 0
\ee
in (\ref{etaparagen}), where $\delta(a)$ is the Dirac function. This choice gives the simplest separable solution of the differential equation (\ref{etap}). Therefore, we have 
\begin{equation}\label{etapara}
   \eta(X,Y) = c \sqrt{(X^2+b)(Y^2-b)} ,
\end{equation}
which is easily shown to satisfy condition (\ref{condeta}) in the form
\be\label{condetapara}
\frac{\eta_{X Y}}{\eta_{X} \eta_{Y} } = \frac{1}{\eta}.
\ee
With solution (\ref{etapara}), equations (\ref{xe1}--\ref{xe2}) are readily solved for $\xi$
\begin{equation}\label{xipara}
   \xi(X,Y) = \frac{c}{4} 
   \arctan{\sqrt{\frac{X^2+b}{Y^2-b}}},
\end{equation}
so that, from (\ref{solTomegaXY}), the conformal vector field is
\be
\widetilde \Omega =
\frac{bc}{8} \frac{X^2+Y^2}{\left((X^2+b)(Y^2-b)\right)^{3/2}}.
\ee
>From this, eq.(\ref{comega2XY}) gives
\be
\Omega = \frac{\widetilde \Omega}{4(X^2+Y^2)}
             = \frac{c^{4} b}{32} \frac{1}{\eta^{3}}.
             \ee
This result suggests to put
\be
c = 2
\ee
in order to simplify formulas and we have the physical vector field  from
\be\label{omegapara}
2 \Omega = \frac{b}{\eta^{3}}, \quad 
   \eta(X,Y) = 2 \sqrt{(X^2+b)(Y^2-b)}.
             \ee
An important consequence of this result is that, together with $W$, $\Omega$ too depends on the coordinates only through $\eta$. This implies that the right hand of eq.(\ref{ntcond5}) for $W$ vanishes, so that, taking into account (\ref{condetapara}), we get simply
\be\label{ntcond5para} 
W''(\eta) + \frac{3}{\eta} W'(\eta) = 0.
\ee
The solution for the scalar potential is then
\be\label{Wparasol} 
W = \frac{\beta}{{\eta}^{2}} 
\ee
and we are actually led to a situation analogous to that encountered above in the example examined in polar coordinates. It is interesting to remark that potential (\ref{Wparasol}) is separable if considered in the purely scalar situation, since, using (\ref{etapara}), we get
\be
W = \frac{\beta}{4(X^2+Y^2)}
       \left(\frac{1}{X^2+b} + \frac{1}{Y^2-b}\right).
       \ee
Reverting to Cartesian coordinates through
\be 
X = \sqrt{\frac{r+x}{2}}, \quad
Y = \sqrt{\frac{r-x}{2}},\ee
the scalar and vector potentials are respectively given by
 \be\label{Wparacart} 
W = \frac{\beta}{y^{2}-4b(x+b)} 
\ee
and
\be\label{omegaparacart} 
\Omega = \frac{b}{2 \left(y^{2}-4b(x+b)\right)^{3/2}}. 
\ee
The second invariant using parabolic coordinates is
 \be\label{quad2RTpara}  
\begin{split} I^{(2)} = &\frac{1}{X^2+Y^2} 
              \left[ Y^2 \left(X' \right)^2 - 
                      X^2 \left(Y' \right)^2 \right] + \\
            & \frac{Y (X^{2} + b) X' -
                       X (Y^{2} - b) Y'}
             {(X^2+Y^2) \sqrt{(X^2+b)(Y^2-b)}} + \\
             &\frac{b + 8 \beta (Y^2-X^2-b)}{4(X^2+b)(Y^2-b)} ,    \\ \end{split}       \ee
where the $\tau$ time variable is used and, therefore, it is conserved along the solution of the transformed equations of motion of the form
(\ref{cwmotion}). Using relations (\ref{veltr1}--\ref{veltr2}) between velocities in the two gauges, 
\ba
X' &=& \sqrt{2(r+x)} {\dot x} + \sqrt{2(r-x)} {\dot y}, \\
Y' &=& \sqrt{2(r+x)} {\dot y} - \sqrt{2(r-x)}  {\dot x} ,
\ea
we can transform the invariant into Cartesian coordinates
\be\label{quad2RTparaCart}  
I^{(2)} = (y {\dot x} - x {\dot y}) {\dot y}            
             +\frac{y {\dot x} + 2 b {\dot y}}
             {2 \sqrt{y^{2}-4b(x+b) }} + 
             \frac{b - 8 \beta (x+b)}{4(y^{2}-4b(x+b))} .           \ee

\subsection{Elliptical coordinates}

It turns out that even in elliptical coordinates it is possible to find non trivial solutions with a structure closely related to that seen in the examples detailed above in polar and parabolic coordinates. In addition, in elliptical coordinates it also exists a solution with a constant vector field (constant `angular velocity'), whereas this possibility appears to be absent in parabolic coordinates. 

A common feature of all these cases is that both the scalar and vector potentials, $W$ and $\Omega$, can be expressed only in terms of $\eta$. This implies that the right hand of eq.(\ref{ntcond5}) for $W$ vanishes, so that, taking into account (\ref{condeta}), this common feature is embodied in the equation
\be 
W''(\eta) + \Phi(\eta) W'(\eta) = 0,
\ee
with suitable $\Phi$. Analogously to what seen in the polar cases, other solutions may very well exist but they are not so easy to find.

According to (\ref{SAB24}), we now have
\be A = 
\Delta^2 \sinh^2 {X} , \quad B =  \Delta^2  \cos^2 Y. \ee 
In this case eq.(\ref{xe1}) becomes
\be\label{etaell}
\sinh {X} \cosh {X}  \eta_{Y} - \sin Y \cos Y \eta_{X} - (\sinh^2 {X} + \cos^2 Y) \eta_{XY} = 0.
\ee
A simple solution of this PDE is the following
\be\label{etaell1}
\eta(X,Y) = a \Delta^2 (\sinh^2 {X} - \cos^2 Y) +
b \Delta^4 (\sinh^2 {X} + \cos^2 Y)^{2},
\ee
with $a$ and $b$ constants. Using this solution, equations (\ref{xe1}--\ref{xe2}) give for $\xi$
\begin{equation}\label{xiell1}
   \xi(X,Y) = a \ln \left[ \Delta^2 (\sinh^2 {X} + \cos^2 Y) \right]+ 
   \frac12 b \Delta^2 (\cosh {2X} - \cos {2Y}).
\end{equation}
Computing the conformal vector potential from (\ref{solTomegaXY}), we see that the first solution is trivial since its Laplacian vanishes giving a null vector field. We remark that the analogous phenomenon also occurs in the parabolic coordinates case, where it is possible to find several additional solutions of equation (\ref{etap}) generating a vanishing vector field.

The second solution appearing in (\ref{xiell1}) instead gives 
\be
\widetilde \Omega =
b \Delta^2 (\cosh {2X} + \cos {2Y}) =
2 b \Delta^2 (\sinh^2 {X} + \cos^2 Y).
\ee
We get that the conformal field is proportional to the conformal factor $|S|$ so that, from (\ref{comega2XY}), we have that the vector field is constant
\be
\Omega = 2 b \equiv \Omega_{0}. \ee
Condition (\ref{condeta}) for the non trivial solution (\ref{etaell1}) with $a=0$ is
\be\label{condetaell1}
\frac{\eta_{X Y}}{\eta_{X} \eta_{Y} } = \frac{1}{2\eta}
\ee
so that eq.(\ref{ntcond5}) for $W$ is
\be 
W''(\eta) + \frac{3}{2\eta} W'(\eta) = 0,
\ee
whose solution is
\be
W = \frac{\alpha}{\sqrt{\eta}},
\ee
or, using again (\ref{etaell1}),
\be
W (X,Y) = \frac{\beta}{|S|} =
\frac{\beta}{\Delta^2 (\sinh^2 {X} + \cos^2 Y)},
\ee
where $\alpha$ and $\beta$ are arbitrary constants. This solution is the simplest separable scalar potential in elliptical coordinates, which is therefore shown to be integrable also in a uniformly rotating system: this result is already known for its application in the modelling of rotating galaxies \cite{vg}. The second invariant (\ref{quad2RT}) using elliptical coordinates is
\begin{eqnarray*}
I^{(2)} &=& \frac{1}{\sinh^2 {X} + \cos^2 Y} 
              \left[ \cos^2 Y (X')^2 - 
                      \sinh^2 (Y')^2 \right]
   + \frac12 \Omega_{0} (\sin 2 Y      X' - \sinh 2 X  Y')  \\ &-& 
             \beta \frac{\sinh^2 {X} - \cos^2 Y}{\sinh^2 {X} + \cos^2 Y} 
             - \Omega_{0}^{2} (\sinh^2 {X} + \cos^2 Y)^{2}.           
             \end{eqnarray*}
Reverting to Cartesian coordinates through
\ba
2\Delta^2 \sinh^2 X &=& r^2 - \Delta^2
             + \sqrt{(r^2+\Delta^2)^2-4\Delta^2 y^2} ,\label{elcart1} \\
         2\Delta^2 \sin^2 Y &=& r^2 + \Delta^2
             - \sqrt{(r^2+\Delta^2)^2-4\Delta^2 y^2}, \label{elcart2} \ea
the scalar potential becomes
\be
W (x,y) = \frac{\beta}{\sqrt{r^{4}+2 \Delta^2 (x^{2} - y^{2}) + \Delta^4}}
\ee
and, using relations (\ref{veltr1}--\ref{veltr2}) between velocities in the two gauges,
we can transform the invariant into Cartesian coordinates
\be\label{quad2RTellCart}  
\begin{split}
I^{(2)} = &(y {\dot x} - x {\dot y})^{2}           
             - \Delta^2 {\dot x}^{2}
             + (\Delta^2 - r^{2})\left(2 \Omega_{0} y {\dot x} - W(x,y) \right) 
             + 2 \Omega_{0} (\Delta^2 + r^{2}) x {\dot y} + \\
            & \Omega_{0}^{2} (r^{4}+2 \Delta^2 (x^{2} - y^{2}) + \Delta^4).   \\        \end{split}
             \ee

We pass now to investigate a more complex class of systems. A general solution of equation (\ref{etaell}) is
\begin{equation}\label{etaellgen}
  \begin{split}
   \eta(X,Y) = 
   &\int F(a) \sqrt{(\Delta^2 \sinh^2 {X}+a)(a-\Delta^2  \cos^2 Y)} d a + \\
  & \int G(a) \frac{(\Delta^2 \sinh^2 {X} + \Delta^2  \cos^2 Y)^{2}}
  {\left((\Delta^2 \sinh^2 {X}+a)(a-\Delta^2  \cos^2 Y)\right)^{3/2}} d a, \\ \end{split}
\end{equation}
where $a$ is an arbitrary real parameter and $F$ and $G$ two arbitrary smooth functions. Comparing this solution with that of (\ref{etaparagen}) we can guess analogous developements. Therefore we try with the simple solution
\be
F = \delta(a-b), \quad G = 0,
\ee
which corresponds to the simplest separable solution of the differential equation (\ref{etaell}). Therefore, we have 
\begin{equation}\label{etaells}
   \eta(X,Y) = \sqrt{(\Delta^2 \sinh^2 {X}+b)(b-\Delta^2  \cos^2 Y)} ,
\end{equation}
which satisfies condition (\ref{condeta}) in the same form as in
(\ref{condetapara}). With solution (\ref{etaells}), equations (\ref{xe1}--\ref{xe2}) give the following expression for $\xi$
\begin{equation}\label{xiell}
   \xi(X,Y) = - {\rm arctanh}
   {\sqrt{\frac{b+\Delta^2 \sinh^2 {X}}{b-\Delta^2  \cos^2 Y}}},
\end{equation}
so that, from (\ref{solTomegaXY}), the conformal vector field is
\be
\widetilde \Omega =
\frac{b \Delta^2 (b - \Delta^2) 
        (\sinh^2 {X} + \cos^2 Y)}
        {\left((\Delta^2 \sinh^2 {X}+b)(b-\Delta^2  \cos^2 Y)\right)^{3/2}}.
\ee
>From this, eq.(\ref{comega2XY}) gives
\be
\Omega = \frac{\widetilde \Omega}
                        {\Delta^2 (\sinh^2 {X} + \cos^2 Y)}
             = \frac{b (b - \Delta^2)}{2 \eta^{3}}.
             \ee
As remarked above, from this results it follows that the right hand of eq.(\ref{ntcond5}) for $W$ vanishes, so that, taking into account (\ref{condetapara}), we get again (\ref{ntcond5para}) so that the solution for the scalar potential is 
\be\label{Wellsol} 
W = \frac{\beta}{{\eta}^{2}} =
\frac{\beta}{(\Delta^2 \sinh^2 {X}+b)(b-\Delta^2  \cos^2 Y)}. 
\ee
Even potential (\ref{Wellsol}) is separable if considered in the purely scalar situation, since it can be written in the form
\be
W = \frac{\beta}{\Delta^2 (\sinh^2 {X} + \cos^2 Y)}
       \left(\frac{1}{b-\Delta^2  \cos^2 Y} - 
              \frac{1}{\Delta^2 \sinh^2 {X}+b}\right).
       \ee
Using the explicit coordinate transformation (\ref{elcart1}--\ref{elcart2}), the scalar and vector potentials are respectively given by
 \be\label{Wellcart} 
W = \frac{\beta}{b y^{2}+(b - \Delta^2)(x^{2}+b)} 
\ee
and
\be\label{omegaellcart} 
\Omega = \frac{b (b - \Delta^2)}
              {2 (b y^{2}+(b - \Delta^2)(x^{2}+b))^{3/2}}. 
\ee
Finally, the second invariant using elliptical coordinates is
\begin{eqnarray*}
I^{(2)} &=& \frac{\cos^2 Y (X')^2 - \sinh^2 (Y')^2}
                          {\sinh^2 {X} + \cos^2 Y} \\ &+&   
                 \frac{1}{\sinh^2 {X} + \cos^2 Y}   \left(      \sin 2 Y {\sqrt{\frac{b+\Delta^2 \sinh^2 {X}}{b-\Delta^2  \cos^2 Y}}}
     X' - \sinh 2 X                
       {\sqrt{\frac{b-\Delta^2  \cos^2 Y}{b+\Delta^2 \sinh^2 X}}}                                
       Y' \right) \\ &+& 
             \frac{b (\Delta^2-b)+2 \beta (\Delta^2(\cos^2 Y-\sinh^2 X)-b)}
             {(\Delta^2 \sinh^2 {X}+b)(b-\Delta^2  \cos^2 Y)} ,          
             \end{eqnarray*}
whereas using Cartesian coordinates is given by
\be\label{quad2RTellCart2}  
I^{(2)} = (y {\dot x} - x {\dot y})^{2}           
             - \Delta^2 {\dot x}^{2}
             +2\frac{b y {\dot x} + (\Delta^{2} - b) x {\dot y}}
             {\sqrt{b y^{2}+(b - \Delta^2)(x^{2}+b)}} +
             \frac{b(b-\Delta^{2}) + 2 \beta(b - \Delta^2 + x^{2}+y^{2})}
             {b y^{2}+(b - \Delta^2)(x^{2}+b)}.           \ee

%%%%
\section{Concluding remarks}

We have investigated Hamiltonian systems with vector potentials admitting a second invariant which is linear or quadratic in the momenta. In our approach, weak and strong invariants are treated in a unified setting where the strong invariants emerge as special cases. As for scalar potentials, the integrable systems can be greatly simplified by introducing certain standardized coordinates, as given in (\ref{Scases1}-\ref{Scases4}). It is a striking result that these standardized coordinate systems for systems with strong invariants exactly coincide with the classical separable coordinates for scalar potentials.

This work is an extension and improvement of the approach to integrable vector potential Hamiltonians which was proposed in \cite{gp}. However, there still remain issues which need clarification. In particular, it should be possible to obtain a better understanding of the integrability conditions, especially the role of the condition (\ref{condeta}) for the structure of the strongly invariant case.

\vfill\eject

\end{document}